\documentclass[a4paper]{article}

\usepackage{INTERSPEECH2021}
\usepackage{booktabs}
\usepackage{tabularx}
\usepackage{url}
\usepackage{multirow,bigdelim}
\usepackage{pbox}
\usepackage{hyperref}

\hypersetup{
    colorlinks=true,
    citecolor=blue,
    linkcolor=blue,
    filecolor=magenta,      
    urlcolor=blue,
}

\title{INTERSPEECH 2022 Audio Deep Packet Loss Concealment Challenge}
\name{Lorenz Diener, Sten Sootla, Solomiya Branets, Ando Saabas, Robert Aichner, Ross Cutler}

\address{Microsoft Corporation}
\email{lorenzdiener@microsoft.com}

\newcommand{\etal}{~et~al.}
\newcommand{\ca}{\raisebox{0.5ex}{\texttildelow}}

\begin{document}

\maketitle
\begin{abstract}
Audio Packet Loss Concealment (PLC) is the hiding of gaps in audio streams caused by data transmission failures in packet switched networks. This is a common problem, and of increasing importance as end-to-end VoIP telephony and teleconference systems become the default and ever more widely used form of communication in business as well as in personal usage.
  
This paper presents the INTERSPEECH 2022 Audio Deep Packet Loss Concealment challenge. We first give an overview of the PLC problem, and introduce some classical approaches to PLC as well as recent work. We then present the open source dataset released as part of this challenge as well as the evaluation methods and metrics used to determine the winner. We also briefly introduce PLCMOS, a novel data-driven metric that can be used to quickly evaluate the performance PLC systems. Finally, we present the results of the INTERSPEECH 2022 Audio Deep PLC Challenge, and provide a summary of important takeaways.

\end{abstract}
\noindent\textbf{Index Terms}: packet loss concealment, speech enhancement, real-time processing

\section{Introduction}

With the transition of digital voice communication away from circuit switched analog telephony to end-to-end digital packet-switched telephone systems -- voice-over-IP (VoIP) telephony -- the issue of how to best process and transmit speech over packet networks has become increasingly important. One important aspect of this is how to handle situations where packet transmission breaks down because packets are lost entirely (packet loss) or arrive too late to be used in a real time communication setting (loss due to excessive jitter). Ideally, we would like to build systems that are robust to a small number of lost packets, and degrade gracefully for longer loss bursts. This process of restoring parts of the signal that are missing in a way that is minimally disturbing to the users of a system is called Packet Loss Concealment (PLC).

Classically, PLC is treated as part of the design of the compression / coding and decompression / decoding algorithms (codecs) used to transmit speech, operating in the coding algorithms feature space. While these techniques allow for basic smoothing over of losses, they can sometimes struggle even with short losses: A study by Sun\etal~\cite{sun_impact_2001} found that for codecs used in mobile telephony, any packet loss during a voiced speech segment causes almost the entirety of that segment to be degraded. Techniques that can be used independently of the base PLC approach, such as forward error correction -- transmitting parts of frames multiple times in anticipation of a loss -- can help for short losses, but put additional strain on a network connection that is already straining to transmit all packets. Lost packets manifest as dropouts or audible distortion to the user -- indeed, ``distorted audio'' is one of the top problem tokens in Microsoft Teams post-call quality surveys.

A less studied but very promising approach that, due to advances in hardware and algorithms, is performing PLC via machine learning and neural networks -- Deep PLC. By leveraging audio and call data to build models that can predict future frames in a data-driven manner and without the strong assumptions classical techniques require, it may be possible to transparently hide shorter losses almost completely, and smooth over longer loss bursts in a less distracting way than current PLC implementations are able to. 

An issue that complicates current Deep PLC research is that as it stands, there are no standard benchmark datasets and evaluation methods, so comparing approaches is extremely limited. With the INTERSPEECH 2022 Deep PLC challenge, we hope to rectify this situation by introducing a realistic evaluation dataset that draws packet loss patterns from actual calls, a methodology for evaluation that will facilitate the comparison of different approaches, and a new objective metric that we hope will help researchers to iterate on their approaches.

In the following paper, we will review classical PLC implementations, contrast them with some recent work in Deep PLC, and finally, present the dataset, evaluation methods and results from the INTERSPEECH 2022 Deep PLC Challenge.

\section{Background and related work}

\subsection{Classical approaches}
Classically, PLC is performed in the feature space of the codec used to packetize, encode, and decode speech data, preceding the decoding step. This is, of course, especially necessary if the codec in question relies on information from consecutive packages in decoding. The techniques used for this have been iteratively improved with newer generations of codecs, but have in essence remained the same from the original work done for the Global System for Mobile Telecommunication~\cite{hellwig_speech_1989} to the modernized Adaptive Multi-Rate Wideband codec used in UMTS~\cite{3rd_generation_partnership_project_adaptive_2004} and the EVS codec used in Voice-over-LTE~\cite{lecomte_packet-loss_2015}. 

The basic principle of these approaches is to continue decoding as if the change in the coded speech parameters from the last known good frames continues according to some expert-crafted fixed prediction function (linearly, in the simplest case), with gradual attenuation and replacement of the signal with comfort noise. More modern codecs improve upon this scheme by classifying missing frames into different types (e.g., silence, voiced speech, non-periodic, ...) and using different prediction schemes for different frame types.

An additional technique employed in many codecs on top of these methods is forward error correction: When bad network conditions are detected, the sender can transmit redundant information about past frames so that short losses can be better compensated for if the next frame after the loss is already available. The downside of this technique is that it introduces network overhead and additional latency. For a deeper survey on past proposed schemes, refer to  Thirunavukkarasu\etal~\cite{thirunavukkarasu_survey_2015}.

\subsection{Machine learning approaches}

\subsubsection{Audio inpainting}
The field of \emph{audio inpainting} concerns itself with filling gaps in an audio signal, with recent approaches largely favoring statistical methods. This is related to the PLC problem, though in a relaxed form: It is often assumed that the entire signal, other than the parts that need to be in-painted, is available, and computation time is not generally considered.

A simple statistical approach is presented by Rodbro\etal~\cite{rodbro_hidden_2006}. They propose to use a Hidden Markov Model on the pitch, gain, and spectral envelope of packets that can then be used to either directly predict future frames, directly fill gaps between frames, or be used as a frame type predictor to choose a prediction scheme.

Bahat\etal~\cite{bahat_self-content-based_2015} present a dictionary based scheme: They learn a dictionary of audio blocks and continue an audio sequence by finding the best matching audio block to continue the sequence with. To find the best match, they use both a Markov model as well as the feature space distance between the start of the block and the last known good part of the audio sequence. The dictionary is created on the fly from correctly transmitted audio. This is a simple and effective way to leverage audio data to fill audio gaps, though computationally expensive and not easily adaptable beyond a single speaker.

Neural network approaches that consider the whole spectrogram are presented by Kegler~\etal~\cite{kegler_deep_2020}  and Nair~\etal~\cite{nair_cascaded_2021}. The former treats the task of audio inpainting as a vision task: They train a U-Net~\cite{ronneberger_u-net_2015} model to map masked spectrograms (magnitude and phase angle), optionally with an additional channel indicating whether a part of the spectrum was masked or not, to the unmasked spectrum, using a perceptual VGG network-based loss. They report an improvement over a baseline linear predictive coding approach for most cases. The latter use a joint time-frequency approach that also performs general speech enhancement: They use a time-domain U-Net to fill gaps and then use a frequency domain U-Net on the magnitude output of the first network to remove distortions (while keeping the phase from the time-domain U-Net signal).

Others approach the problem from a multi-modal perspective: Both Zhou\etal~\cite{zhou_vision-infused_2019} and Morrone\etal~\cite{morrone_audio-visual_2021} present approaches where an available video feed of the speaker is used to assist in the restoration of lost audio segments, using a convolutional neural network with adversarial loss and a recurrent neural network approach respectively. While approaches such as these are outside the purview of current deep PLC research, they may become relevant in the context of video telephony, where in  future techniques, available video frames may help reconstruct the audio stream and vice-versa.

\subsubsection{Real-time deep packet loss concealment}
In recent years, both hardware and algorithms have advanced dramatically, allowing neural network-based approaches to enter practical use in many areas of speech processing. This is also happening for PLC. Here, we give an overview of recent work targeting real-time, causal usage.

Stimbert\etal present an approach based on WaveRNN~\cite{kalchbrenner_efficient_2018}: They condition a WaveRNN on the recent time domain history as well as, through a convolutional conditioning network operating in the frequency domain, more long term history. Unlike other methods presented in this section, this network outputs samples autoregressively, one by one, instead of outputting full blocks of audio data at a time. They present methods for how their network can be used for real-time inference, operating on the jitter buffer of the NetEQ codec used in the Google Duo telecommunication software~\cite{noauthor_improving_nodate}, and state that their implementation of this method significantly improved user satisfaction metrics.

One issue when training neural networks for audio generation is what loss to use. Generative adversarial networks provide a framework within which a loss does not need to be defined, but can be co-learned by training a discriminator network that tries to classify a sample as either being generated or sampled from the training set, and then training the generation network to try to fool the discriminator. GAN-based approaches have been shown to be able to efficiently generate audio waveforms~\cite{kong_hifi-gan_2020}. A basic initial PLC approach based on GANs is presented by Shi\etal~\cite{shi_speech_2019}. They train a convolutional encoder-decoder network operating on time domain audio blocks. They report comparable quality in terms of several objective metrics (PESQ~\cite{rix_perceptual_2001}, STOI~\cite{taal_algorithm_2011}, SNR) compared to a frequency-domain deep neural network even when this baseline has perfect phase information available and only needs to predict magnitude. Pascual\etal~\cite{pascual_adversarial_2021} present a GAN-based approach where the generator input is the Mel-spectrogram of the available signal, and the output is the time-domain continuation of this signal. They show that this method improves upon several baselines (including some of the systems used in real codecs, such as NetEQ and ITU G722.1) in terms of Mel-Cepstral Distortion~\cite{kubichek_mel-cepstral_1993} as well as SESQUA~\cite{serra_sesqa_2021}. Finally, Wang\etal~\cite{wang_temporal-spectral_2021} present a GAN-based system with a fully time-domain U-Net style convolutional generator and mixed time/frequency domain discriminator, which allows their adversarial loss to both capture fine short-term details in the waveform as well as long-term relationships in the spectrum. They evaluate in terms of PESQ, STOI, and SNR and are able to show an improvement over several strong baselines -- even over non-causal models when their own model is being used causally. 

Other methods show that Deep PLC systems can be trained even with simpler losses. Lin\etal~\cite{lin_time-domain_2021} present a convolutional-recurrent model performing next frame prediction in the time domain, trained to minimize the mean absolute error, with or without look-ahead. In addition to the usual metrics, they also measure speech recognizer word error rate, and are able to show that Deep PLC has the potential to improve this metric as well. Similarly, Mohamed\etal~\cite{mohamed_concealnet_2020} present a recurrent neural network for packet loss concealment in the context of far-end emotion recognition. They show that using such a network before their emotion classifier improves its accuracy, further illustrating the potential of Deep PLC to improve the performance of downstream speech processing tasks.

\section{Challenge description}

The objective of the INTERSPEECH 2022 Audio Deep Packet Loss Concealment challenge is to fill regions where audio has been lost due to packet loss in such a way as to ideally hide these losses from listeners and to maximize intelligibility. 

Note that effects from de- and encoding or buffering and time stretching are not considered as part of this challenge.

\subsection{Dataset and evaluation}

The dataset provided for the INTERSPEECH 2022 Audio Deep Packet Loss Concealment challenge consists of a set of audio and meta-data files, divided into three splits: A train split, validation split and blind test split. The training and validation splits contain, for each included audio sample:

\begin{itemize}
    \item A clean audio file
    \item A ``lossy" audio file with packet loss regions zeroed out
    \item A text file with loss metadata
\end{itemize}

The metadata indicates, with one line for each 20 millisecond segment of audio, whether the packet has been lost (line contains a ``1") or not (line contains a ``0"). The ``lossy" audio files have the corresponding segments zeroed out. Additionally, the dataset includes, for each split, a file with file-level metadata about the included clips in csv format. The training set is intended as a quick-start set for training PLC models, though participants were free to use any other datasets to improve their models. The blind set release did not include clean audio.

The training split includes a total of 23184 clips, while the validation and blind test split each include 966 clips.

Training and validation splits were released to participants on January 19th, 2022. The blind split data was released on March 1st, 2022, after which participants had one week to generate and submit results. Blind split reference data was released after the conclusion of the challenge. The dataset is now available as a fully open source dataset for use in further evaluations by any interested researcher\footnote{\url{https://github.com/microsoft/PLC-Challenge}}.

\subsubsection{Dataset construction}

The dataset was constructed by a stratified sampling of actual packet loss traces observed in calls made by Microsoft Teams users, applied to randomly chosen segments of audio from a podcast dataset. 

\begin{figure}[ht]
    \centering
    \def\svgwidth{0.95\columnwidth}
    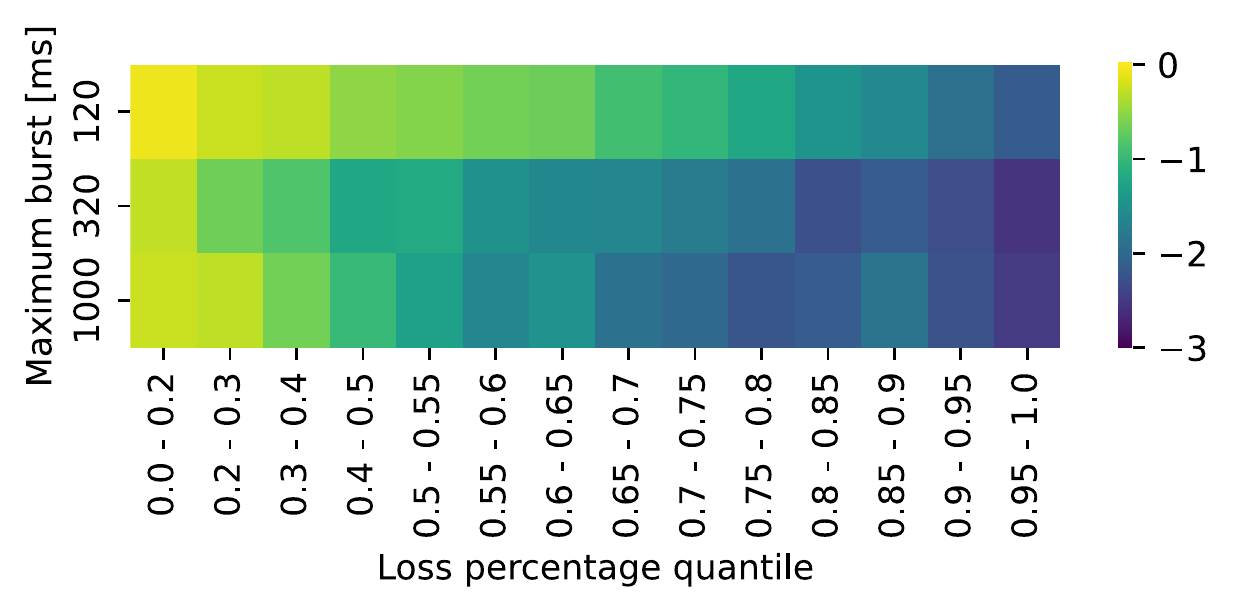\vspace{-8pt}
    \caption{CMOS for lossy audio files for different maximum burst sizes and loss percentage quantiles included in our dataset.}
    \label{fig:lossy_cmos}
\end{figure}

The traces were sampled as follows: First, 10 second segments with at least one lost packet were randomly extracted from packet loss traces from Teams calls. These segments were then divided into three subsets according to the maximum burst loss length in the trace:

\begin{itemize}
    \item up to 120 milliseconds (504 blind set clips)
    \item between 120 and 320 milliseconds (308 blind set clips)
    \item between 320 and 1000 milliseconds (154 blind set clips)
\end{itemize}

Traces with burst losses longer than 1000 milliseconds were discarded. Each subset was then divided into 14 cells according to packet percentage loss quantiles. Finally, an equal amount of traces was sampled from each cell (with more traces being sampled for the subsets with shorter maximum burst losses).

To create audio with packet losses, clips of audio of approximately 10 seconds of length were sampled from a base public domain podcast dataset (LibriVox Community Podcasts), cutting in low volume regions to not split up words. The losses from the sampled traces were then applied to the resulting audio files by zeroing out the corresponding regions in the audio clips.

Figure~\ref{fig:lossy_cmos} shows CMOS scores (compare section~\ref{sec:eval} for the different burst subsets and packet loss fraction percentiles in the blind set. For low burst lengths, a smooth, relatively linear relationship between loss and CMOS is visible. For the longer burst lengths, the gradient is shifted substantially towards worse scores (with a lot of noise for the very high burst lengths likely owing to the relatively small amount of clips per cell).

\subsubsection{Latency and compute restrictions}

In practice, PLC is used as part of an audio telecommunications system. This means that Deep PLC models have tight latency requirements. In the 2022 INTERSPEECH Deep PLC Challenge, the requirements given for models were as follows:

\begin{itemize}
    \item The PLC must take less than the stride time to process a frame on an Intel Core i5 quad-core machine clocked at 2.4 GHz or equivalent processors.
    \item  Total algorithmic latency including the frame size, stride time, and any look-ahead must be $\leq$ 20ms.
\end{itemize}

\subsubsection{Evaluation} \label{sec:eval}
The evaluation criterion is split into two parts, weighted equally: Crowd-Sourced \emph{Comparative Mean Opinion Score} (CMOS), and \emph{Word Accuracy} (WAcc). The Mean Opinion Score is calculated by having raters rate clips using a crowd-sourcing framework~\cite{naderi_open_2020}, aiming for five ratings for each clip in the blind set. Since in PLC, it is possible to have clips where a system may fill a long gap with silence where in the original, speech was present, raters are given the reference for every file, and asked to rate the given file compared to this reference on a 7 point categorical scale ranging from -3 to 3 -- this is called comparative category rating. For details on evaluation methodology, please refer to our previous work.~\cite{naderi_open_2020}

The word accuracy is calculated using Microsoft Azure Cognitive Services speech recognition by subtracting the returned \emph{Word Error Rate} (WER) from 1. The score ranges from 0 to 1, where 1 is a perfect score.

To calculate the final score, we normalize the CMOS range from 0 (the same as the reference) to -3 (much worse than the reference) to a range of 0 to 1 such that higher is better, and take the mean of the normalized CMOS and the word accuracy. The final score by which we rank submissions is calculated as:

\begin{equation*}
    \text{score} = \frac{\left(1 + \frac{\text{CMOS}}{3}\right) + \left(1 - \text{WER}\right)}{2}
\end{equation*}



\subsection{PLCMOS}

\begin{table*}[ht]
\centering
\caption{2022 INTERSPEECH Audio Deep PLC Challenge Results. Teams marked as ``tied": Difference not significant ($p \geq 0.05$). Parameter count and structure given where known (TD: Time-Domain, FD: Frequency Domain).}
\begin{tabular}{@{}rlrrrrrrr@{}}
\toprule
 Rank &                                  Team ID &  PLCMOS &  DNSMOS &   CMOS &  WAcc &  Final score & \#Parameters & Approach \\
\midrule
             &     Clean (reference) &   4.51 &   3.89 & 0     &   0.97 &  0.98 &     -      & - \\ 
    1        &              \#12 &   4.28 &   3.80 & -0.55 &   0.88 &  0.85 & \ca2.36M  & TD U-Net \\
    2        &                \#1 &   3.74 &   3.79 & -0.64 &   0.88 &  0.84 & \ca5.9M   & RNN + LPCNet \\
    3 (tied) &          \#9 &   3.83 &   3.68 & -0.81 &   0.87 &  0.80 &           & TD Encoder-Decoder \\
    3 (tied) &  \#14 &   3.98 &   3.69 & -0.84 &   0.86 &  0.79 & \ca4.97M  & TD Sequence-to-One \\
    5        &                  \#11 &   3.28 &   3.51 & -1.10 &   0.86 &  0.75 & \ca36k    & FD Conv. Net \\
    6        &                  \#10 &   3.48 &   3.74 & -1.04 &   0.83 &  0.74 &           & \\
             & Zero-filling baseline &   2.90 &   3.44 & -1.23 &   0.86 &  0.73 &      -     & - \\
    7        &    \#6 &   2.90 &   3.48 & -1.31 &   0.86 &  0.71 & \ca3.77M  & \\
\bottomrule
\end{tabular} \vspace{-6pt}
\label{tab:results}
\end{table*}

\begin{table}[ht]
\centering
\caption{2022 INTERSPEECH Audio Deep PLC Challenge: Metrics for non-eligible models.}
\begin{tabular}{@{}lrrrr@{}}
\toprule
               Team  ID &  PLCMOS &  DNSMOS &   CMOS &  WAcc. \\
\midrule
           \#2 &   4.406 &   3.958 & -0.279 &          0.854 \\
\#13 &   3.274 &   3.453 & -1.115 &          0.853 \\
               \#3 &   3.371 &   3.293 & -1.284 &          0.852 \\
\bottomrule
\end{tabular} \vspace{-8pt}
\label{tab:results_ineligible}
\end{table}

To help participants evaluate their systems, we developed and provided access to PLCMOS\footnote{\url{https://github.com/microsoft/PLC-Challenge/tree/main/PLCMOS}}, a neural network model trained to estimate the score a human rater would assign to a given audio file. There are multiple models provided: A semi-intrusive model that requires a reference, but does not require the reference to be aligned with the lossy signal, and a non-intrusive model that functions entirely without reference audio. The model employs a convolutional-recurrent encoder structure and is trained on ground truth labels produced by P.808 ACR labelling~\cite{naderi_open_2020}. It produces a score between 1 and 5 as an estimate of human mean opinion scores. We report PLCMOS scores for our participants in Table~\ref{tab:results}. Overall, on the data evaluated for this challenge, PLCMOS and subjective CMOS showed a Pearson correlation coefficient of \ca0.95 and a Spearman rank correlation coefficient of \ca0.93, compared to a PCC of \ca0.89 and SRCC of \ca0.90 for DNSMOS~\cite{reddy_dnsmos_2021}, which hasn't been trained on packet loss data. We plan to further improve this system in the future with more data, including the data gathered during this challenge.

\section{Results}

We present the results of our evaluation in Table~\ref{tab:results}, including PLCMOS and DNSMOS~\cite{reddy_dnsmos_2021} metrics. In addition to participant results, we also list the metrics for a zero-filling baseline, which simply replaced the losses with silence -- the outputs of this baseline comprises the ``lossy" dataset made available to the contestants.

CMOS results were obtained using the Amazon Mechanical Turk service and \cite{naderi_open_2020}, targeting 5 ratings per clip. A total of 421 raters participated in the evaluation. An initial run had a rejected answer rate of \ca30\% due to failed trapping clips or raters submitting more than the allowed amount of ratings. Rejected ratings were re-ran. After removing raters with low acceptance rates ($<$90\%) and with few completed rating tasks ($<$3), the final results presented here are based on at least 4 votes per clip for 75\% of clips submitted for rating, with a confidence interval of \ca0.035 for the CMOS score.

Word accuracy was calculated based on 944 clips of the blind set. 22 clips containing multi-lingual or non-speech audio were excluded from WER evaluation because the speech recognition service used could not reliably produce transcriptions of these clips. Ground truth transcriptions were created using Azure Cognitive Services to transcribe the clean audio files, followed by human inspection and correction where needed.

We performed statistical tests to get a better understanding of how significant the differences between models are. After establishing that significant differences exist using ANOVA testing, we performed post-hoc pairwise t-tests for the CMOS votes, using the same procedure as in our previous challenges. We find that, while most differences are significant at a level of $p < 0.05$, we cannot establish significant differences between the Alibaba inc. and Oldenburg University teams. In our final results, we denote these teams as tied.

In addition to the models presented in Table~\ref{tab:results}, some models submitted for the challenge used a larger algorithmic latency than intended due to oversights in modeling. The prevalence of this issue underscores the need to carefully verify constraints when building models for use in real-time and low-latency applications. The metrics for these submissions are listed in Table~\ref{tab:results_ineligible} -- while we could not consider them for this challenge, they are nevertheless interesting, especially in regard to potential future challenges going towards a full neural jitter buffer, where a larger look-ahead will in some cases be appropriate.

\section{Conclusion}

This paper has presented the results of the INTERSPEECH 2022 Audio Deep PLC Challenge. Looking back, we believe this challenge has managed to achieve its intended goal -- to stimulate research into neural network-based PLC and to bring researchers interested in this topic together. Some key takeaways include:

\begin{itemize}
    \item While imperfect, objective metrics are still quite predictive of the final results. We'll continue improving PLCMOS to facilitate the development of ever-better PLC systems.
    \item Scores in the challenge were quite close, especially towards the top end, and significant differences could not always be established. For future challenges, we will look into constructing a more difficult task, possibly going into the direction of a full neural jitter buffer so that models can infill and modify the time scale with which samples are played back, within some given maximum latency.
    \item One effect we did not expect is the relative prevalence of accidental leaks of future information beyond the allowed maximum algorithmic latency. For future challenges, we will try to equip participants with tools they can use to easily verify that their model fits the constraints provided.
\end{itemize}

We would like to thank all participants for their efforts in creating and submitting models, and INTERSPEECH for hosting the Audio Deep Packet Loss Concealment Challenge.

\bibliographystyle{IEEEtran}
\bibliography{interspeech_plc_challenge_2022}

\end{document}